\documentclass{iopart}
\usepackage{amssymb}
\usepackage{iopams}
\usepackage{bm}
\usepackage{psfrag}

\usepackage{epsf,graphicx,color}

\newcommand{\eff}{{\rm eff}}

\newcommand{\la}{\langle}
\newcommand{\ra}{\rangle}

\begin{document}

\letter{The decay of photoexcited quantum systems: 
a description within the statistical scattering model}

\author{T Gorin$^1$, B Mehlig$^2$, and W Ihra$^1$}

\address{\hspace*{-1.5mm}\mbox{}$^1$Theoretische Quantendynamik, Physikalisches 
     Institut, Universit\" at Freiburg, Hermann-Herder-Str. 3,
     D-79104 Freiburg, Germany}
\address{\hspace*{-1.5mm}\mbox{}$^2$Physics 
and Engineering Physics, Gothenburg University/Chalmers Gothenburg, Sweden}

\begin{abstract}
The decay of photoexcited quantum systems
(examples are photodissociation of molecules and autoionization of
atoms) can be viewed as a half-collision process (an incoming
photon excites the system which subsequently decays by
dissociation or autoionization). For this reason, the
standard statistical approach to quantum scattering,
originally
developed to describe nuclear compound reactions,
is not directly applicable. 
Using an alternative approach,
correlations and fluctuations of observables
characterizing this process were first derived in
[Fyodorov YV and Alhassid Y 1998 {\it Phys. Rev.} A {\bf 58}  R3375].
Here we show how the results cited above,
and more recent results incorporating direct decay processes,
can be obtained from the standard statistical scattering
approach by introducing one additional channel.
\end{abstract}


\submitto{\JPA}

\ead{Thomas.Gorin@physik.uni-freiburg.de}

\maketitle

\begin{figure}[t]
  \psfrag{i}{n}
 \centerline{\input{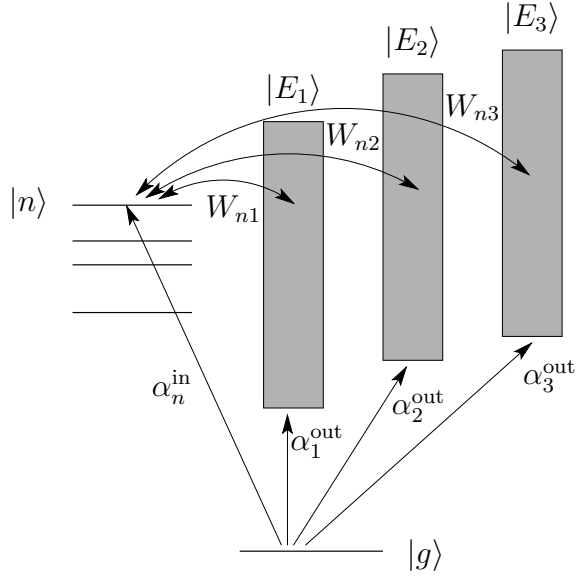}}
 \caption{\label{fig:fig1}
    Direct photoionization from an initial state 
    $|g\rangle$ into the subspace of unbound states
    ($M$ continuum channels $|E_a\rangle$, here: $M=3$) 
    with transition amplitudes $\alpha^{\rm out}_a$, and 
    indirect photoionization
    via $N$ bound states $|n\rangle$ with transition 
    amplitudes $\alpha^{\rm in}_n$. 
    The states $|n\rangle$ turn into  resonances due to
    couplings to the continua with matrix elements $W_{na}$.
    %In the figure the coupling matrix elements between the $P$-space and
    %the $Q$-space are expressed in the eigenbasis of the $P$-space.
    }
\end{figure}

Molecular photodissociation~\cite{kir00,dob96,dob95b,pes97,pes95,rei96}
and atomic autoionization~\cite{zak95} are examples of
quantum-mechanical
decay processes:  
excited quantum systems decay into one or several scattering continua
(decay channels). The situation is illustrated 
schematically in figure~\ref{fig:fig1}.  
An atom or a molecule is excited from its ground state 
or a low lying state $|g\ra$ 
by a photon. Photoabsorption either results in direct excitation of the decay 
channels $|E_a\rangle$ (where $a=1,\ldots,M$)
or in excitation of quasi-bound states $|n\ra$  (where $n=1,\ldots,N$). 
Decay into a %continuum 
channel $|E_a\rangle$ then proceeds via residual interactions
between the quasi-bound states and the continua.

Correlations and fluctuations of the quasi-bound states
determine the statistical properties of the indirect decay
cross section. If the corresponding classical dynamics is
sufficiently chaotic, these properties can be modeled by
random matrix theory~\cite{haa91,meh91,GMW98}. In 
reference~\cite{Fyod98} this approach was adopted to characterize
the distribution and two-point correlations of the indirect photodissociation
cross-section, in the absence of direct decay channels. 
The results of \cite{Fyod98} were obtained
by calculating the statistical properties of a resolvent of an effective
non-Hermitian Hamiltonian. Equivalently they may be obtained from
eigenvector correlations \cite{sav97,meh98,fyo02} 
of the non-Hermitian effective Hamiltonian.
In the limit of weak continuum coupling,
cross-section correlations can be calculated
using perturbation theory~\cite{Alha98}.
The results of reference~\cite{Alha98} have been generalized
in~\cite{Alha03}
where interference effects 
between direct and indirect decay paths in the photoabsorption 
cross section were studied.
%For instance, in the case of isolated resonances, this
%interference gives rise to so-called Beutler-Fano profiles \cite{Fano61},
%characterized by the Fano parameter. In~\cite{Ihra02} the statistics of that 
%parameter has been studied for the case of chaotic dynamics in the subspace
%of quasi-bound states.

How are the results obtained in \cite{Fyod98,Alha98,Alha03}
related to those derived for nuclear and mesoscopic scattering
processes within the so-called statistical scattering model
(also termed the Heidelberg 
model~\cite{Mahaux69,VWZ85,FyoSom97,GMW98})? 
Since photodecay  is a half-collision process~\cite{Rost98,Schinke93},
the statistical scattering model as such
is not directly applicable. In order to establish a connection to the 
statistical scattering model, it has been suggested~\cite{AlhSel:priv} 
to introduce an additional channel representing photoexcitation. 
In this paper we show how to describe the decay
of photoexcited quantum systems using this idea. Our results
can be summarized as follows:  We demonstrate that partial
and total decay cross sections can be obtained from an extended
(almost) unitary scattering matrix by introducing an additional channel 
for photoexcitation and a linear transformation reminiscent of
the ``Engelbrecht-Weidenm\"uller" transformation~\cite{EngWei73}.
We use that relation to obtain a non-perturbative
result for the two-point correlation function 
of the total cross section in a random-matrix model,
which allows for arbitrary coupling strengths between
quasi-bound states, decay channels and the photoexcited initial state.
%In particular, it allows to study the effects of interfering direct and 
%indirect decay paths in detail.

%Let $|g\rangle$ be the ground state of the system in question,
%or a low lying state from which 
%photoexcitation takes place, 
Let $|\alpha\rangle = 
\hat \mu|g\rangle$ 
be the ground state dressed with the dipole operator $\hat \mu$. 
The partial photodissociation (photoionization) cross section into channel 
$a$ is proportional to \cite{Schinke93,Fri98}
\begin{equation}
  \sigma_a(E) \propto \left|\la \alpha|\Psi^{(a)}(E)\ra\right|^2\,,
\label{eq:part_cross}\end{equation}
where the scattering wave function $|\Psi^{(a)}\rangle$ is outgoing in 
channel $a$ only. The total photodissociation cross section is given by
\begin{equation}
\sigma(E) = \sum_{a=1}^M \sigma_a(E)
\end{equation} 
($M$ is the number of open channels).
We follow the approach adopted in \cite{Fyod98,Alha03} 
where resonant and direct  processes were
treated separately
by means of the Feshbach approach \cite{Fesh58}:
the Hilbert space of excited states is divided into 
a subspace of bound states, and its complement containing 
all unbound states. The vector $|\alpha\rangle$ is
decomposed accordingly $|\alpha\rangle= |\alpha^{\rm in}\rangle\oplus
|\alpha^{\rm out}\rangle$.
Choosing the bound space
appropriately, the scattering problem projected onto the complement looses 
all resonant features. 
The Hamilton operator
in the bound space is represented by an $N\times N$ random matrix
$\bi{H}_0$, with eigenstates $|n\rangle$, $n=1,\ldots N$.
The $N$ dipole transition matrix elements
to $|n\rangle$ are denoted by $\alpha^{\rm in}_n$,
and $\bm{\alpha}^{\rm in} 
= (\alpha^{\rm in}_1 ,\,\ldots ,\,\alpha^{\rm in}_N)^T$ .
Correspondingly, the $M$ 
(energy-independent) dipole 
transition matrix elements from $|g\rangle$
to the decay channels
are  
$\alpha^{\rm out}_a$, and 
$\bm{\alpha}^{\rm out} = 
(\alpha^{\rm out}_1 ,\,\ldots ,\, \alpha^{\rm out}_M)^T$.
The matrix elements of the residual interaction
$\hat W$ are denoted by $W_{na} = \langle n|\hat W|E_a\rangle$,
and  $\bm{W}_a= (W_{1a} ,\,\ldots ,\, W_{Na})^T$
are the column vectors of $\bm{W}$.
For the total cross section, the following result has been obtained
in \cite{Alha03}:
\begin{equation}
\fl 
\sigma(E) 
= \|\bm{\alpha}^{\rm out}\|^2 - \pi^{-1} {\rm Im}\left[\,
   (\bm{\alpha}^{\rm in} + 
   \rmi\pi\; \bm{W}\; \bm{\alpha}^{\rm out})^\dagger\;  
   \frac{1}{E-\bm{H}_{\rm eff}}\;
   (\bm{\alpha}^{\rm in} 
   - \rmi\pi\; \bm{W}\; \bm{\alpha}^{\rm out})\, \right] \; .
   \label{eq:sigabs}
\end{equation}
Here $\bm{H}_{\rm eff} = \bm{H}_0 - {\rm i}\pi \bm{W} 
\bm{W}^\dagger$ is an effective non-Hermitian Hamiltonian.
%and
%the column vectors 
%$\bm{W}_a= (W_{1a} ,\,\ldots ,\, W_{Na})^T$ contain the coupling matrix 
%elements between the basis states $|n\rangle$
%of the bound space and the scattering states 
%$|E_a\ra$ (Figure~\ref{fig:fig1}).
%depicts the different roles of these 
%parameters, where the basis states $|n\ra$ have been chosen so that $H_0$ 
%is diagonal. 
The partial cross section is given by
\begin{equation}
\sigma_a(E) = \left| \alpha^{\rm out}_a  + \bm{W}_a^\dagger\; 
 \frac{1}{E-\bm{H}_{\rm eff}}
   \; (\, \bm{\alpha}^{\rm in} - \rmi\pi\; \bm{W}\; \bm{\alpha}^{\rm out}\, )\, 
   \right|^2 \,.
\label{eq:sigc}
\end{equation}
It can be shown by purely algebraic manipulations that, indeed, 
$\sum_{a=1}^M \sigma_a(E) = \sigma(e)$.
%The column vectors 
%$\bm{W}_a= (W_{1a} ,\,\ldots ,\, W_{Na})^T$ contain the coupling matrix 
%elements between the basis states $|n\rangle$
%of the bound space and the scattering states 
%$|E_a\ra$ (Figure~\ref{fig:fig1}).

We now prove that the cross sections given in
(\ref{eq:sigabs}) and (\ref{eq:sigc}) can be expressed in terms of 
the extended $(1+M)\times (1+M)$ unitary scattering matrix
\begin{equation}
\fl \bm{S}(E,\delta)= \mathbf{1}_{1+M} - 2\rmi\pi\; 
         \bm{V}^\dagger\; \frac{1}{E-  \bm{F}_{\rm eff}}\; \bm{V}
         \quad\mbox{with}
\quad
\bm{V} =\big (  \bm{\alpha}^{\rm in}\, \delta/2\pi\,,
         \bm{W} \big) \; ,
\label{F_Smat}\end{equation}
and $\bm{F}_\eff = \bm{H}_0 - \rmi\pi\, \bm{V}\, \bm{V}^\dagger$.
This maps our problem onto the statistical scattering model.
% thereby mapping our problem onto the statistical scattering model.
% In (\ref{F_Smat}),  
% $\bm{F}_\eff = \bm{H}_0 - \rmi\pi\, \bm{V}\, \bm{V}^\dagger$. 
It is 
convenient to enumerate the channels from $0$ to $M$, so that the new 
channel, referring to photoexcitation, gets the index 0. The
corresponding column vector of $\bm{V}$ is scaled with a real positive
parameter $\delta$ (which will be taken to zero at the end of the calculation). 
Now, consider
transform $\bm{S}'(E,\delta)= \bm{u}(\delta)\, \bm{S}(E,\delta)\, 
\bm{v}^T(\delta)$,
where the matrices $\bm{u}(\delta)$ and $\bm{v}(\delta)$ are given by
\begin{equation}\label{eq:S_matrix_trafo}
\fl \bm{u}(\delta) = \left( \begin{array}{c|c}
%     \rmi & \bar\alpha^*_1\,\delta/2 \ldots \bar\alpha^*_M\,\delta/2 \\ \hline
     \rmi & ({\bm{\alpha}^{\rm out}})^\dagger\,\,\delta/2 \\ \hline
              \begin{array}{c} 0      \\
                               \vdots \\
                               0 \end{array} &
              \begin{array}{ccc}   &        &   \\
                                   &  \mathbf{1}_{M} &   \\
                                   &        &  \end{array}
              \end{array} \right)\,, \qquad
\bm{v}(\delta) = \left( \begin{array}{c|c}
%         \rmi & \bar\alpha_1\,\delta/2 \ldots \bar\alpha_M\,\delta/2 \\ \hline
     \rmi & ({\bm{\alpha}^{\rm out}})^T\,\,\delta/2 \\ \hline
              \begin{array}{c} 0      \\
                               \vdots \\
                               0 \end{array} &
              \begin{array}{ccc}   &        &   \\
                                   & \mathbf{1}_M &   \\
                                   &        &  \end{array}
              \end{array} \right) \; .
\end{equation}
The transformation (\ref{eq:S_matrix_trafo}) is reminiscent of an 
``Engelbrecht-Weidenm\" uller'' transformation \cite{EngWei73},
although for $\delta > 0$, $\bm{u}(\delta)$ 
and $\bm{v}(\delta)$ are not unitary.
We now show that the cross sections (\ref{eq:sigabs}) and
(\ref{eq:sigc}) can be expressed in terms of
the elements of
$\bm{S}'(E,\delta)$ as follows
\begin{eqnarray}
\sigma_a(E) &={\phantom 2\,}  \lim_{\delta\to 0} \delta^{-2}
   \left|\frac{\delta}{2}\; \alpha^{\rm out}_a + S'_{a0}(E,\delta) \right|^2 
\label{F_sigc}\\
\sigma(E) &= 2\, \lim_{\delta\to 0} \delta^{-2}\left[ \frac{\delta^2}{4}\; 
   \|\bm{\alpha}^{\rm out} \|^2 + 1 + {\rm Re}\; S'_{00}(E,\delta) \right] \; .
\label{F_sigabs}\end{eqnarray}
In order to derive these relations, 
we  compute the corresponding matrix elements of $\bm{S}'(E,\delta)$. Let
$1\le a \le M$:
\begin{eqnarray}
\fl 
S'_{a0}(E,\delta) &= \sum_{b,c=0}^M
  u_{ab}(\delta)\; S_{bc}(E,\delta)\; v_{0c}(\delta)
= \rmi\; S_{a0}(E,\delta) + \sum_{c=1}^M S_{ac}(E,\delta)\; 
   \frac{\delta}{2}\; \alpha^{\rm out}_c \nonumber\\ 
\fl &= \frac{\delta}{2}\; \alpha^{\rm out}_a + \delta\; \bm{W}_a^\dagger\; 
   \frac{1}{E- \bm{F}_\eff}\; (\bm{\alpha}^{\rm in} 
   -\rmi\pi\; \bm{W}\; \bm{\alpha}^{\rm out}) 
\label{comp1}\\
\fl 
S'_{00}(E,\delta) &= - S_{00}(E,\delta) + \rmi\;\frac{\delta}{2}
   \sum_{b=1}^M 
   \left[ (\alpha^{\rm out}_b)^*\; S_{b0}(E,\delta) + S_{0b}(E,\delta)\; 
   \alpha^{\rm out}_b \right] \nonumber\\
\fl&\hspace*{0.92cm}   + \frac{\delta^2}{4}\sum_{b,c=1}^M 
   (\alpha^{\rm out}_b)^*\; 
   S_{bc}(E,\delta)\; \alpha^{\rm out}_c \nonumber\\
\fl &= -1 +\rmi\;\frac{\delta^2}{2\pi}\; (\bm{\alpha}^{\rm in})^\dagger\; 
   \frac{1}{E-\bm{F}_\eff}\; \bm{\alpha}^{\rm in} \nonumber\\
   \fl &\qquad+ 
    \frac{\delta^2}{2}\left[ 
   (\bm{\alpha}^{\rm out})^\dagger\; \bm{W}^\dagger\; 
   \frac{1}{E- \bm{F}_\eff}\; \bm{\alpha}^{\rm in} + 
   (\bm{\alpha}^{\rm in})^\dagger\; \frac{1}{E- \bm{F}_\eff}\; \bm{W}\; 
   \bm{\alpha}^{\rm out} \right] 
\label{comp2}\\
%\fl &\qquad
%   + \frac{\delta^2}{4}\left( \|\bm{\alpha}^{\rm out}\|^2 -2\pi\rmi\; 
%   (\bm{\alpha}^{\rm out})^\dagger\; \bm{W}^\dagger\; 
%     \frac{1}{E- \bm{F}_\eff}\; \bm{W}\; 
%   \bm{\alpha}^{\rm out}\right) \nonumber\\
\fl &= -1 + \frac{\delta^2}{4}\; \|\bm{\alpha}^{\rm out}\|^2 + 
   \frac{\rmi\, \delta^2}{2\pi}\; 
   (\bm{\alpha}^{\rm in} + \rmi\pi\; \bm{W}\; \bm{\alpha}^{\rm
   out})^\dagger \; 
   \frac{1}{E- \bm{F}_\eff}\; 
   (\bm{\alpha}^{\rm in} - \rmi\pi\; \bm{W}\; \bm{\alpha}^{\rm out}) \; .
\nonumber\end{eqnarray}
Since $\bm{F}_\eff \to \bm{H}_\eff$ in the limit $\delta\to 0$,
we obtain expressions~(\ref{eq:sigabs}) and (\ref{eq:sigc}) for
the cross sections. This is easily verified by inserting~(\ref{comp1}) 
and~(\ref{comp2}) into the left hand side of~(\ref{F_sigc}) 
and~(\ref{F_sigabs}), respectively.

Equations~(\ref{F_sigc}) and~(\ref{F_sigabs}) enable us to compute the 
average total cross section, as well as its correlation function, defined as
\begin{equation}
C(E,w) = \la \sigma(E-w\, \Delta/2)\, \sigma(E+w\, \Delta/2) \ra -
   \la \sigma(E)\ra^2 \,.
\label{eq:C}\end{equation}
Here $\Delta$ is the average local distance between neighboring resonances,
and $\la\ldots\ra$ denotes an energy average, often replaced by an ensemble 
average in theory. For simplicity, we assume that the column vectors of 
$\bm{V}$ in equation~(\ref{F_Smat}) are pairwise orthogonal
%. Note that one can 
%account for non-orthogonal column vectors with the help of an additional 
%Engelbrecht-Weidenm\" uller transformation on $S'(E,\delta)$.
(see below).
In this case, the average extended scattering matrix 
$\la S(E,\delta)\ra$ is real and diagonal, with diagonal elements given 
by~\cite{Gor02}
\begin{equation}
\fl \la S_{aa}(E,\delta)\ra = \frac{1-\lambda_a}{1+\lambda_a}\,, \qquad
\lambda_a = \pi^2\, \varrho_0\, \|W^{(a)}\|^2\,,\quad \mbox{and}\quad
\lambda_0 = \frac{\delta^2}{4}\; \varrho_0\, \|\bm{\alpha}^{\rm in}\|^2 \; ,
\end{equation}
where $N\varrho_0 = 1/\Delta$ is the level density.
Starting from equation~(\ref{F_sigabs}) and assuming
that $\bm{H}_0$ corresponds to a classically chaotic 
system, we obtain the expression for the average total cross section
derived in \cite{Alha03}:
\begin{eqnarray}
\fl \la\sigma(E)\ra = \frac{\|\bm{\alpha}^{\rm out} \|^2}{2} + 
     \lim_{\delta\to 0} \; \frac{2}{\delta^2}\left[\rule{0pt}{17pt} \right.
      1+ \sum_{a,b=0}^M {\rm Re}\left[\rule{0pt}{11pt} u_{0a}\, v_{0b}\, 
     \la S_{ab}(E,\delta)\ra\right] \left. \rule{0pt}{17pt} \right] \nonumber\\
= \|\bm{\alpha}^{\rm out} \|^2 + \varrho_0\, \|\bm{\alpha}^{\rm in}\|^2
     - \sum_{a=1}^M |\alpha_a^{\rm out}|^2\, \frac{\lambda_a}{1+\lambda_a} \; .
\label{av-sigabs}\end{eqnarray}
In a similar way, we may write for the autocorrelation function of $\sigma(E)$
\begin{eqnarray}
\fl  C(E,w) 
  = \lim_{\delta\to 0}\; \frac{2}{\delta^4}\; \sum_{a,b=0}^M\sum_{c,d=0}^M 
   u_{0a}(\delta)\, u_{0c}(\delta)\, v_{0b}(\delta)\, v_{0d}(\delta)\;
\nonumber\\
 \times\; {\rm Re}\left[\rule{0pt}{11pt}
     \la S_{ab}(E-w\Delta/2)\,S_{cd}^*(E+w\Delta/2)\ra - \la S_{ab}(E)\ra\; 
     \la S_{cd}^*(E)\ra\right] \; .
\label{eq:16}\end{eqnarray}
It has been used that the correlation function of two $S$-matrix 
elements is different from zero
only if one $S$-matrix element is complex conjugated while the other is not
(see reference~\cite{Gor02} for details). Equation~(\ref{eq:16}) expresses
the autocorrelation function of the total absorption cross section in terms
of correlation functions of elements of the extended scattering matrix, defined 
in~(\ref{F_Smat}). 
Usually, the energy dependence of the autocorrelation function is mainly due to
the energy dependence of the coupling parameters
$\bm{\alpha}^{\rm out}$, $\|\bm{\alpha}^{\rm in}\|^2$ and
$\lambda_1,\ldots,\lambda_M$. This dependence should be weak in 
order to allow energy averaging over a sufficiently large window.
In cases where the corresponding S-matrix correlation
functions are available, equation~(\ref{eq:16}) enables us to obtain the 
autocorrelation function of the total cross section in closed form.
\begin{figure}[b]
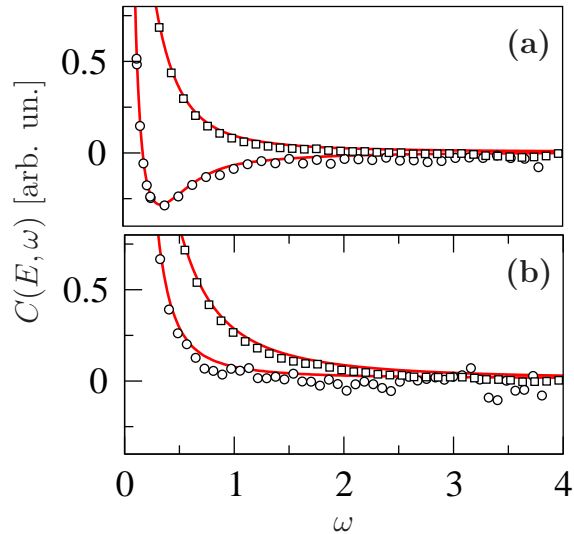

  \psfrag{a}{\hspace*{-2mm}\large {\bf (a)}}
    \psfrag{b}{\hspace*{-2mm}\large {\bf (b)}}
      \psfrag{XLABEL}{}
        \psfrag{YLABEL}{}
          \centerline{\includegraphics[clip,width=0.55\columnwidth]{ca.eps}}
            \psfrag{YLABEL}{\hspace*{0.9cm}\raisebox{1mm}{\large
            $C(E,\omega)$ [arb. un.]}}
              \psfrag{XLABEL}{\hspace*{0.5cm}\large $\omega$}
                \centerline{\hspace*{1.5mm}\includegraphics[clip,width=0.5575\columnwidth]{cb.eps}}
                 \caption{\label{fig:fig2} Correlation function
                 (\ref{eq:C}) of
                  the total cross section, time-reversal invariant case,
                  results according
                   to equation~(15) (lines) and from the numerical
                   diagonalisation
                    of random matrices with $N=128$ (symbols).
                     \textbf{(a)}: without direct coupling
                     ($\bm{\alpha}^{\rm out}=0$),
                      $M=2$, $\gamma=0.02$ ($\circ$),  and
                       $\gamma=0.5$ ($\Box$). \textbf{(b)}: with direct
                       coupling,
                        $M=2$, $\gamma=0.02$, $\alpha^{\rm out}_j = 2.5$
                        ($\circ$), and
                         $\gamma=0.5$, $\alpha^{\rm out}_j = 0.5$
                         ($\square$).  } \end{figure}

As an illustration, let us assume time-reversal invariance 
for the collisional system and predominantly chaotic dynamics in the large 
($N\rightarrow\infty$) bound space. Then, the required $S$-matrix correlation 
functions are given by the Verbaarschot-Weidenm\" uller-Zirnbauer (VWZ) 
integral~\cite{VWZ85}.
Using a Fourier representation developed in~\cite{Gor02} we obtain:
\begin{eqnarray}
\fl C(E,w) = 4\int\!\!\rmd t\; \rme^{-2\pi\rmi\, w t}
 \int_{\hspace*{-1cm}\raisebox{-2mm}{$\scriptstyle{\rm max}(0,t-1)$}}^t\!\!\!\!\!\!%
 \rmd r\int_0^r \!\!\rmd u\;
     \frac{(t-r)(r+1-t)}{(2u+1) (t^2-r^2+x)^2}\; \prod_{e=1}^M
     \frac{1-T_e(t-r)}{(1+2T_e r + T_e^2 x)^{1/2}} \nonumber\\[2mm]
\fl\qquad\times \left\{  
     \left(\Delta_0 - \frac{1}{4}\sum_{a=1}^M \tau_a\; \sqrt{1-T_a}\; 
     \Delta_a\right)^2
 + \Pi_{00} + \sum_{a=1}^M \frac{\tau_a}{2} \; \Pi_{0a} + \sum_{a,b=1}^M 
     \frac{\tau_a\, \tau_b}{16}\; \Pi_{ab} \right\} \; .
\end{eqnarray}
where $x= u^2\, (2r+1)/(2u+1)$, and where 
$\tau_a = T_a\, |\alpha_a^{\rm out}|^2/(\varrho_0\, \|\bm{\alpha}^{\rm in}\|^2)$.
As usual, it is more convenient to work with the transmission coefficients
$T_a = 4\lambda_a/(1+\lambda_a)^2$ to describe the coupling between the subspace
of quasi-bound states and the decay channels. The functions $\Delta_a$ and 
$\Pi_{ab}$ are defined as follows:
\begin{eqnarray}
\fl \Delta_a = \frac{r+ T_a\, x}{1+ T_a(2r+T_a x)} + \frac{t-r}{1-T_a(t-r)}
\nonumber\\
\fl \Pi_{ab} = \frac{T_a T_b\, x^2 +[T_a T_b\, r+(T_a+T_b)(r+1)-1] x + r(2r+1)}
   {(1+2T_a r + T_a^2\, x) (1+2T_b r + T_b^2\, x)} \nonumber\\
 + \frac{(t-r)(r+1-t)}{[1-T_a(t-r)][1-T_b(t-r)]} \; .
 \label{eq:DeltaPi}
\end{eqnarray}
In the special case of $M\!=\!1$ open channel, a result corresponding
to (15,16)  was
derived, by a different approach, in \cite{fyo02a}.
In the absence of direct processes (i.e., when all direct dipole
transition amplitudes $\alpha^{\rm out}_a$ are zero) equation~(15) 
reduces
to the result derived in~\cite{Fyod98}.
Note that in (\ref{eq:DeltaPi}), the limit $\delta\to 0$ has 
been taken. This implies that the transmission coefficient corresponding to the
``photo channel'' has to be set to zero.

 In figure~\ref{fig:fig2}, the correlation function $C(E,\omega)$ is
 shown: as obtained from equation~(15), valid in the limit of 
$N\rightarrow\infty$, and as obtained from the
 numerical diagonalisation of finite matrices of size $N=128$. 
 In the presence
 of direct decay channels, finite-size effects 
 are seen to be somewhat
 larger, as compared with purely indirect decay.

We conclude with two remarks. First, the VWZ-integral can also be used 
to calculate the average of the partial photoabsorption cross sections. 
However, for the correlation functions of the partial cross sections no exact 
analytical formulae are known. In this case one must use approximate results, 
such as the ``rescaled'' Breit-Wigner approximation \cite{Alha98,Gor02}.
Second, in theoretical studies of the statistical scattering model, it
is often assumed that the channel vectors (the column
vectors of $\bm{V}$) are
orthogonal to each other: one can always obtain the 
original scattering matrix by an appropriate unitary transformation from a
simpler scattering matrix with orthogonal channel vectors. 
%This
%transformation can be obtained from the singular value decomposition of 
%$\bm{W}$. 

\ack
Financial support from 
the DFG in SFB 276, from the EU Human Potential Programme under contract 
HPRN-CT-2000-00156, and from Vetenskapsr\aa{}det is gratefully acknowledged.

\section*{References}

\bibliographystyle{unsrt}

\end{document}